\documentclass{article}[12pt]
\usepackage{arxiv}

\usepackage[utf8]{inputenc} 
\usepackage[T1]{fontenc}    
\usepackage{hyperref}       
\usepackage{url}            
\usepackage{booktabs}       
\usepackage{amsfonts}       
\usepackage{nicefrac}       
\usepackage{microtype}      
\usepackage{lipsum}
\usepackage{fancyhdr}       
\usepackage{graphicx}       
\usepackage{amsmath}
\graphicspath{{media/}}     

\title{Rotational Dynamics Induced by Low-Energy Binary Collisions of Quantum Droplets
}

\author{
  J. E. Alba-Arroyo, S. F. Caballero-Benitez,{R. J\'auregui *} \\
  Departamento de F\'{\i}sica Cu\'antica y Fot\'onica, \\
 Laboratorio de Simulaciones Computacionales para Sistemas Cu\'anticos (LSCSC)\\ Laboratorio Nacional de Materia Cuántica (LANMAC) ,\\
 Instituto de F\'{\i}sica, Universidad Nacional Aut\'onoma de M\'exico, {Ciudad de M\'exico} {C.P. 04510},\\
  \texttt{*rocio@fisica.unam.mx} \\
  }

\begin{document}
\maketitle

\begin{abstract}
A theoretical analysis of the rotational dynamics induced by off-axis binary collisions of quantum droplets constituted by ultracold atoms is reported.  We focus on quantum droplets formed by  degenerate dilute Bose gases made from binary mixtures of alkaline atoms under feasible experimental conditions. The stability of the ground state is known to be longer for the chosen heteronuclear gases than for the homonuclear ones. In both cases, we find that the dynamics seem to privilege high similarity of the density of each atomic species. However, the evolution of the phase of the corresponding order parameter  differs significantly for heteronuclear admixtures. We evaluate the fidelity as a figure of merit for the overlap between the order parameters of each atomic species.  Dynamical evidence of the differences between the phases of the order parameters  is predicted to manifest in their corresponding linear and  angular momenta. We numerically verify that the total angular and linear momenta are both conserved during  the collision. Some direct correlations between the Weber number and the impact parameter with the distribution of the dynamical variables are established.
\end{abstract}


\section{Introduction}

Self-bound  droplets formed in ultracold atomic Bose gases
were recently predicted \cite{Petrov2015,Petrov2016} and  experimentally observed~\cite{Cabrera2018,Semeghini2018,Derrico2019,Guo2021}.
The stabilization of quantum droplets  in binary mixtures of ultracold atoms arises from a calibrated  competition between interspecies attractive interactions and {intraspecies} repulsive interactions. The~proper description of this regime requires the incorporation of quantum fluctuations, i.e.,~beyond  mean-field effects. They involve either direct extensions of the Lee--Huang--Yang (LHY) formalism~\cite{LHY1957}
to binary mixtures~\cite{Larsen1963} or~ its analogue for dipolar interactions~\cite{Ferrier2016,Schmitt2016,Chomaz2016,Tanzi2019,Bottcher2019}.

Theoretical studies of the rotational phenomena in quantum droplets formed by ultracold atoms  have already been reported both in the 2D regime~\cite{Fetter2009,Malomed2019,Kartashov2019, Luo2021,Li2018,Examilioti2020,Tengstrand2019,Sturmer2021, Kartashov2020, Otajonov2020} and in the 3D regime~\cite{Kartashov2018,Ancilotto2018a, Caldara2022}. In~those studies,
 emphasis is given to the stability---or lack of stability---of optical vortices. As~a consequence, in~general, an~{\it {ansatz}} of an order parameter that exhibits such an interesting feature is proposed  as an initial state.  Then, a dynamical evolution---generated in terms of an extension of the Gross--Pitaevskii equation---is performed. In~this work, we take an alternative point of view: we study the dynamics  derived from off-axis binary collisions of quantum droplets conformed by mixtures of alkaline atoms that were initially in their corresponding ground states. Up~to now, this mechanism has been used to observe the crossover between compressible and incompressible regimes of quantum droplets~\cite{Ferioli2019}.

Frontal collisions of quantum droplets have already been theoretically analyzed in~\cite{Cikojevic2021, Alba2022}. Attention was paid to the survival of the quantum droplets as a function of the relative importance of the inertia of the fluid in terms of the initial kinetic energy and its surface tension, i.e., the~Weber number~\cite{frohn2000, Alba2022}. Here,  we focus on low-energy collisions.  The~initial translational energy is assumed to be controlled by imprinting a specific phase on each otherwise-ground-state droplet; the initial angular momentum requires additional control of the impact~parameter. 

In the following section, the  extended Gross--Pitaevskii equation is written and the general properties of the corresponding ground state are described. An~accurate analytical expression for the latter is provided. The~initial order parameter is introduced, as well as expressions of the corresponding inertial tensor,  translational kinetic energy,  Weber number, and linear and angular momenta.
Then, we report specific results for the evolution of the dynamical properties of the droplets during binary collisions. We compare the general features of homo- and heteronuclear quantum droplets with parameters that are feasible for experimental realization. Finally, we discuss the main qualitative and quantitative~results.    

\section{Evolution~Equation}

Consider a mixture of two ultracold Bose gases with atomic 
masses $m_a$ and $m_b$, densities $n_a$ and $n_b$, and~scattering lengths $\mathrm{a}_{\alpha\beta}$, $\alpha,\beta=a,b$. 
We assume that the intraspecies scattering lengths $\mathrm{a}_{\alpha\alpha}>0$, while the interspecies scattering length $\mathrm{a}_{ab}<0$. 
Under the weak coupling condition $n\mathrm{a}^3\ll 1$, the~long-wavelength instabilities expected from a mean-field theory approach  are cured by quantum fluctuations at shorter wavelengths~\cite{Petrov2015,Petrov2016}.
An expression for those fluctuations was introduced by Lee, Huang, and Yang 
as a first term in a power-series expansion in the parameter $(n\mathrm{a}^3)^{1/2}$ for homogeneous quantum fluids~\cite{LHY1957}, and~it was extended by Larsen for binary mixtures~\cite{Larsen1963}. 
An effective formalism that considers finite-range effects was reported in~\cite{Cikojevic2018}. Looking for a better understanding of the microscopic dynamics, a formalism involving a pairing field has also been developed for homonuclear binary admixtures~\cite{Hu2020}.

 The expression of the extended Gross--Pitaevskii equation (EGPE) for { the effective order parameters of binary Bose--Bose mixtures $\Psi_a$ } used in the present study is~\cite{Derrico2019}
\begin{eqnarray}
 i \hbar \partial_t \Psi_\alpha &=& \Big(-\frac{\hbar^2}{2m_\alpha} \nabla^2 +   g_{\alpha \alpha} |\Psi_\alpha|^2  + g_{\alpha \beta} |\Psi_\beta|^2 \nonumber \\
&+&  \frac{4}{3 \pi^2} \frac{m_\alpha^{3/5} g_{\alpha \alpha}}{ \hbar^3} (m_\alpha^{3/5} g_{\alpha \alpha} |\Psi_\alpha|^2 + m_\beta^{3/5} g_{\beta \beta} |\Psi_\beta|^2 )^{3/2} 
 \Big) \Psi_\alpha,\label{eq:EGPE}
\end{eqnarray} 
where the coupling strength factors are defined by $g_{\alpha\beta} = 2\pi \hbar^2 \mathrm{a}_{\alpha\beta}/m_{\alpha\beta}$, with~$m_{\alpha\beta}=m_\alpha m_\beta/(m_\alpha + m_\beta )$;
{ $\Psi_a$ is normalized to the number of atoms in species $a$,  $N^{(a)}$.} This equation results from the Gross--Pitaevskii energy functional for a Bose mixture when
both the mean-field term and the LHY correction accounting for quantum fluctuations in the local density approximation are included. The~LHY correction involves a dimensionless function~\cite{Petrov2015,Derrico2019} $f(z=m_b/m_a, u=g_{ab}^2/(g_{aa}g_{bb}),x=g_{bb}n_b/(g_{aa}n_{a})$ that depends on the local density for each atomic species $n_{a}$. In~Equation~(\ref{eq:EGPE}),
this function has been properly parametrized following ref.~\cite{Derrico2019} by the simple expression $(1+x(m_2/m_1)^{3/5})^{5/2}$ for $u=1$. The~latter effective interaction has been widely discussed~\cite{Petrov2015,Derrico2019} and properly describes the general features observed in quantum droplet realizations for both homonuclear~\cite{Cabrera2018} and heteronuclear~\cite{Derrico2019} binary Bose--Bose mixtures. For cases in which the relative motion of the components of the droplet can be neglected, e.g., for~ground-state calculations, the~EGPE acquires a simpler structure~\cite{Petrov2015,Kartashov2018}. ent of the separation vector along the
direction of the relative momenta of the droplets, and b is the impact parameter which as
usual has length units.
Here, we work out the evolution of the droplets using Equation~(\ref{eq:EGPE}) since we are especially interested in possible evidence of these otherwise energy-expensive~fluctuations.

The parameters
\begin{equation}
\xi_a =\hbar\sqrt{\frac{3}{2}\frac{\sqrt{g_{bb}}/m_a + \sqrt{g_{aa}}/m_b}{\vert \delta g\vert \sqrt{g_{aa}}n_a^{(0)}}},\quad
\tau_a =\frac{3\hbar}{2}
\frac{\sqrt{g_{aa}} + \sqrt{g_{bb}}}{\vert\delta g\vert \sqrt{g_{aa}}n_a^{(0)}},\quad  \delta g = g_{ab} + \sqrt{g_{aa} g_{bb}} .\end{equation}
are identified as the natural  units of length $\xi_a$, time $\tau_a$, and {$\delta g$ estimates the  degree of balance between the  interspecies attraction coupling ($g_{ab}<0$) and the geometrical average of the intraspecies repulsive coupling}. In~these equations,
\begin{equation} 
n_\alpha^{(0)} = \frac{25\pi}{1024} \frac{1}{(1+ (m_b/m_a)^{3/5} \sqrt{g_{bb}/g_{aa}})^5}\frac{1}{\mathrm{a}_{\alpha\alpha}^3}\frac{\delta g^2}{g_{aa}g_{bb}},\label{eq:n0}
\end{equation}
{ yields the expected equilibrium density for $ \delta g \ll \vert g_{ab}\vert$, $\delta g \ll \sqrt{g_{aa} g_{bb}}$ and a large number of atoms of each species satisfying $N^{(a)}/N^{(b)} =\sqrt{g_{aa}/g_{bb}}$ as obtained 
in Ref.~\cite{Petrov2015}.}
\subsection*{{Ground}~State}

The self-trapping regime requires a minimum number of atoms $N_c$. In~this regime, the ground state 
is expected to exhibit~\cite{Petrov2015} (i) the normalized order parameters { ($\phi^{(\alpha)}= \Psi_\alpha/\sqrt{N^{(\alpha)}}$ )} for each component that can be taken as real and are related by a simple proportionality factor, $\phi_0^{(a)} =\phi_0^{(b)}$; (ii) an asymptotic, $N^{(\alpha)}\rightarrow \infty$, saturation density $n_\alpha^{(0)}$, Equation~(\ref{eq:n0}) (in fact, for~achieving the quantum droplet regime, it is required that 
$n_b^{(0)}/n_a^{(0)} \approx \sqrt{g_{aa}/g_{bb}}$); (iii) a spherical shape with radius $R_0\approx ( 3 N/4\pi n_\alpha^{(0)}\xi^3)^{1/3}\xi$; and (iv) a surface thickness $dR$  of order $\xi$. 
The  compressible regime corresponds to quantum droplets with a radius $R_0$ whose value is similar to that of their surface thickness $dR$.

We have numerically verified~\cite{Alba2022} that a  Boltzmann function
\begin{equation}
\rho_B(r;N) = \frac{A_1}{1 + \mathrm{exp}((r-R_0)N/dR)}
\label{eq:Boltzmann}
\end{equation}
accurately describes the ground-state order parameters obtained numerically from Equation~(\ref{eq:EGPE}). 
The Boltzman expression allows direct evaluation of  the inertial tensor $I_{ij}$:
\begin{eqnarray}
I_{xx}(N^{(\alpha)}) &=& I_{yy}(N^{(\alpha)}) = I_{zz}(N^{(\alpha)}) = \frac{8\pi m^{(\alpha)}}{3}\int_0^\infty \rho_B(r;N^{(\alpha)})r^4 dr\\
I_{ij}&=& 0, i\ne j
\end{eqnarray}
for  each component of a quantum droplet in its ground~state.

A study of low-energy excitation with spherical symmetry also yields values of the surface tension both for the incompressible regime
\begin{equation}\label{eq:energia_ansatz2_esfericas_hetero} 
\sigma^{(\ell)}_{incomp} = -\frac{\hbar^2}{M}
      \frac{ \int dr  \,  (\partial_r \phi^{(a)}_0 )^2  r^{2\ell-2}   }
     {  \int dr  \, \partial_r \rho^{(a)}_0 r^{2\ell+1} }
\end{equation}
 and for the compressible regime
\begin{equation}\label{eq:energia_ansatz1_esfericas_hetero}
\sigma^{(\ell)}_{comp} = -\frac{\hbar^2}{M}
\frac{ \int dr  (\partial_r \phi^{(a)}_0)^2  
\int dr   \partial_r \rho^{(a)}_0 r^{2\ell+1} 
  }{ \left( \int dr \partial_r \rho^{(a)}_0 r^{\ell+2} \right)^2 } 
  \end{equation}
where $\ell$ is the angular momentum about the symmetry axis of the droplet and\linebreak $M = (4\pi/3)(m_am_b/(N^{(a)}m_b + N^{(b)}m_a)$.

\section{Binary Collisions of Quantum~Droplets}

{Frontal binary collisions of quantum droplets have been experimentally reported~\cite{Ferioli2019}. Those experiments  involve two Bose--Einstein condensates confined in a
a crossed dipole trap with a repulsive thin barrier that splits the BEC along
a given horizontal direction. The~internal interactions are tuned so that the quantum droplet regime is achieved. Removing the thin barrier and switching off the radial dipole trap, the~droplets move towards the center of a vertical levitating trap, acquiring an increasing velocity. After~a time interval $\Delta t$, the vertical trap is switched off and the two droplets keep moving towards each other. The~time interval and the frequency of the radial trap determine the relative momenta of the colliding~droplets. 

By modifying the geometry of the initial double-well trap, the impact  parameter of the droplets can also be controlled.
In this section, we study the corresponding low-energy collisions using Equation~(1).}

\subsection*{{Initial}~State}
Consider the collision of quantum droplets that are initially  kicked off as described by a plane wave factor $\vec k_0 = k_0\hat{e}_\parallel $ and~ separated by a vector
\begin{equation}\vec d = 2\vec d_0 + b \hat{e}_\perp = 2d_0 \hat{e}_\parallel + b \hat{e}_\perp\end{equation} 
with $\hat{e}_\parallel\cdot \hat{e}_\perp=0.$  That is, $2d_0$ is the initial component of the separation vector along the direction of the relative momenta of the droplets, and $b$ is the impact parameter which as usual has length units. Explicitly, 
\begin{eqnarray}
\Psi(\vec r, t=0) &=&\Psi_1(\vec r) + \Psi_2(\vec r)  \nonumber\\
&=&\begin{pmatrix}\psi_{a_1}(\vec r + \vec d_0)\\ \psi_{b_1}(\vec r+ \vec d_0)\end{pmatrix} e^{i\vec k_0\cdot \vec r/2}
+\begin{pmatrix}\psi_{a_2}(\vec r - \vec d_0 - b \hat{e}_\perp)\\ \psi_{b_2}(\vec r- \vec d_0- b \hat{e}_\perp)\end{pmatrix} e^{-i\vec k_0\cdot \vec r/2}. \label{eq:initial}
\end{eqnarray}
The parameter $d_0$ is chosen to numerically guarantee a negligible initial overlap of the droplets.
At $t=0$, the mean value
\begin{equation}
\mathcal{K} =-\frac{\hbar^2}{2}\sum_{i=1,2}\int d^3r \Psi_i^\dagger(\vec r)\begin{pmatrix}  \frac{\nabla^2}{m_{a_i}}  & 0\\ 0 & \frac{\nabla^2}{m_{b_i}}  \end{pmatrix} \Psi_i(\vec r),\label{eq:K}
\end{equation}
which is a measure of the initial kinetic energy of the droplets,
can be decomposed as the sum of the translational kinetic energy of each droplet as a whole $\mathcal{K}_{trans}$ and~an  internal kinetic energy of the atoms within the droplets $\mathcal{K}_{int}$. Thus,  $\mathcal{K}\approx \mathcal{K}_{trans} + \mathcal{K}_{int}$, with
\begin{equation}
\begin{aligned}
& \mathcal{K}_{trans}= \frac{\hbar^2}{2}\Big[ \frac{N^{(a_1)}k_0^2}{4m_{a_1}} + \frac{N^{(b_1)}k_0^2}{4m_{b_1}} +
\frac{N^{(a_2)}k_0^2}{4m_{a_2}} + \frac{N^{(b_2)}k_0^2}{4m_{b_2}}  \Big] , \\
&  \mathcal{K}_{int} = -\frac{\hbar^2}{2}\sum_{i=1,2}\sum_{\alpha=a,b}\int d^3r\psi^*_{\alpha_i}(\vec r)\frac{\nabla^2}{m_{\alpha_i}}\psi_{\alpha_i}(\vec r).
\end{aligned}
\label{eq:trasplusint}
\end{equation}

In addition to the impact  parameter $b$, 
a dimensionless quantity suitable for characterizing binary collisions of classical incompressible droplets, is the Weber number $\mathrm{We}_\ell$ \cite{frohn2000}. For~each one of the droplets, it is defined as the ratio between the  translational kinetic energy $\mathcal{K}_{trans}$ before the collision and the surface energy of excitation evaluated in terms of the surface tension of the droplet; in our case,
\begin{equation}
\mathrm{We}_\ell =\frac{\mathcal{K}_{trans}}{R_0^2\sigma_\ell}\label{eq:Weber}
\end{equation}
where $R_0$ is the initial radius of the droplet. 

 For the collision of two identical quantum droplets initially in their ground state internal configuration, this expression becomes
\begin{equation}\label{eq:We1}
\mathrm{We}_\ell
= \frac{8\pi (dR k_0)^2}{3}\Big(\frac{dR}{R_0}\Big)^2 \frac{((\ell +2)!)^2}{(2\ell +1)!}\Big[1-\frac{1}{(1+ e^{R_0/dR})^2}\Big]^{-1}
                      \Big[\frac{(F_{\ell+1}(R_0/dR))^2}{F_{2\ell }(R_0/dR)}\Big] 
\end{equation}
in the incompressible regime, and~\begin{equation}\label{eq:We2}
\mathrm{We}_\ell=\frac{16\pi (dR k_0)^2}{3}\Big(\frac{dR}{R_0}\Big)^2 (\ell(4\ell^2 -1))
                      \frac{F_{2\ell}(R_0/dR)}{F_{2\ell-3}(R_0/dR)+F_{2\ell -4}(R_0/dR)}
\end{equation}
in the compressible regime. Here, $F_s =(1/\Gamma(s+1)\int_0^\infty(dx\,x^s/(1+e^{x-z}))$ denotes the Fermi--Dirac~integrals.

We numerically solve Equation~(\ref{eq:EGPE}) with the initial condition Equation~(\ref{eq:initial}) by applying the time-evolution operator
\begin{eqnarray}
\Psi(\vec r, t + \Delta t) = \exp(-i \mathcal{H} \Delta t / \hbar ) \Psi(\vec r, t + \Delta t),
\end{eqnarray}
where $\mathcal{H}$ is identified with the right side of Equation~(\ref{eq:EGPE}). We implemented a Strang decomposition (Split-Step Method), which is accurate to the second order in the time step $\Delta t$ 
~\cite{Strang1968,Bao2003}
\begin{eqnarray}
 \exp(-i \mathcal{H} \Delta t / \hbar ) \approx
 \exp(-i \Delta t \mathcal{V}(\vec{r}\prime)  / (2\hbar) ) 
 \exp(-i \Delta t \mathcal{\hat{K}}  / \hbar ) 
  \exp(-i \Delta t \mathcal{V}(\vec{r}\prime)  / (2\hbar) ).
\end{eqnarray}
The variable $\mathcal{\hat{K}} $ is  the kinetic energy operator from Equation~(\ref{eq:K}),  
which was evaluated in $k$-space by means of Fast Fourier transforms (FFTs). The~FFT was implemented following~\cite{numerical-recipes}, and the whole algorithm was developed in Fortran. 

\section{Evolution of Dynamical  Variables during the~Collision}

The  evolution of  binary collisions of quantum droplets according to the EGPE equation was numerically~evaluated.

 The order parameters were obtained both for homonuclear and heteronuclear quantum droplets under feasible experimental conditions. 
We report results~for
\begin{itemize}
\item[(i)] Symmetric $N^{(a)}\approx N^{(b)}=N/2$ mixtures of homonuclear $m_a=m_b$ $^{39}$K atoms and scattering lengths compatible with the Feshbach resonances of such atoms: $\mathrm{a}_{aa} = \mathrm{a}_{bb} = 48.57 \mathrm{a}_0$, $\mathrm{a}_{ab}= - 51.86 \mathrm{a}_0$ \cite{Cabrera2018};
\item[(ii)] Mixtures of $^{41}$K and $^{87}$Rb~\cite{Derrico2019} with scattering lengths $\mathrm{a}_{aa} =62.0\mathrm{a}_0$, $\mathrm{a}_{bb} =100.4\mathrm{a}_0$, $\mathrm{a}_{ab} =-82.0\mathrm{a}_0$. The~stability condition $N^{(b)}/N^{(a)} \approx\sqrt{g_{aa}/g_{bb}}$ was imposed. 
\end{itemize}

The number of atoms $N^{(\alpha)}$ was chosen to guarantee the quantum droplet regime and~to involve droplets that  are expected
to exhibit either compressible or incompressible
features. To~avoid confusion, images and graphs in this section
correspond mostly to the same pair---one homonuclear and one heteronuclear---among many studied selections of $N^{(\alpha)}$ for each droplet. A~discussion
about the main generic findings for other values of  $N^{(\alpha)}$
not explicitly shown in the figures is given  at the end of this~section. 

In~\cite{Alba2022}, the~values of the parameters $A_1$, $R_0$, and $dR$ necessary to obtain an expression for the ground-order parameter  using Equation~(\ref{eq:Boltzmann}) are given so that the initial state Equation~(\ref{eq:initial}) has an analytical representation. We used $\hat e_\parallel = (1/\sqrt{3})(1,1,1)$ and $\hat e_\perp = 1/\sqrt{2})(-1,1,0)$ in the numerical~simulations.

\begin{figure}[h]
 \includegraphics[width=0.95\textwidth]{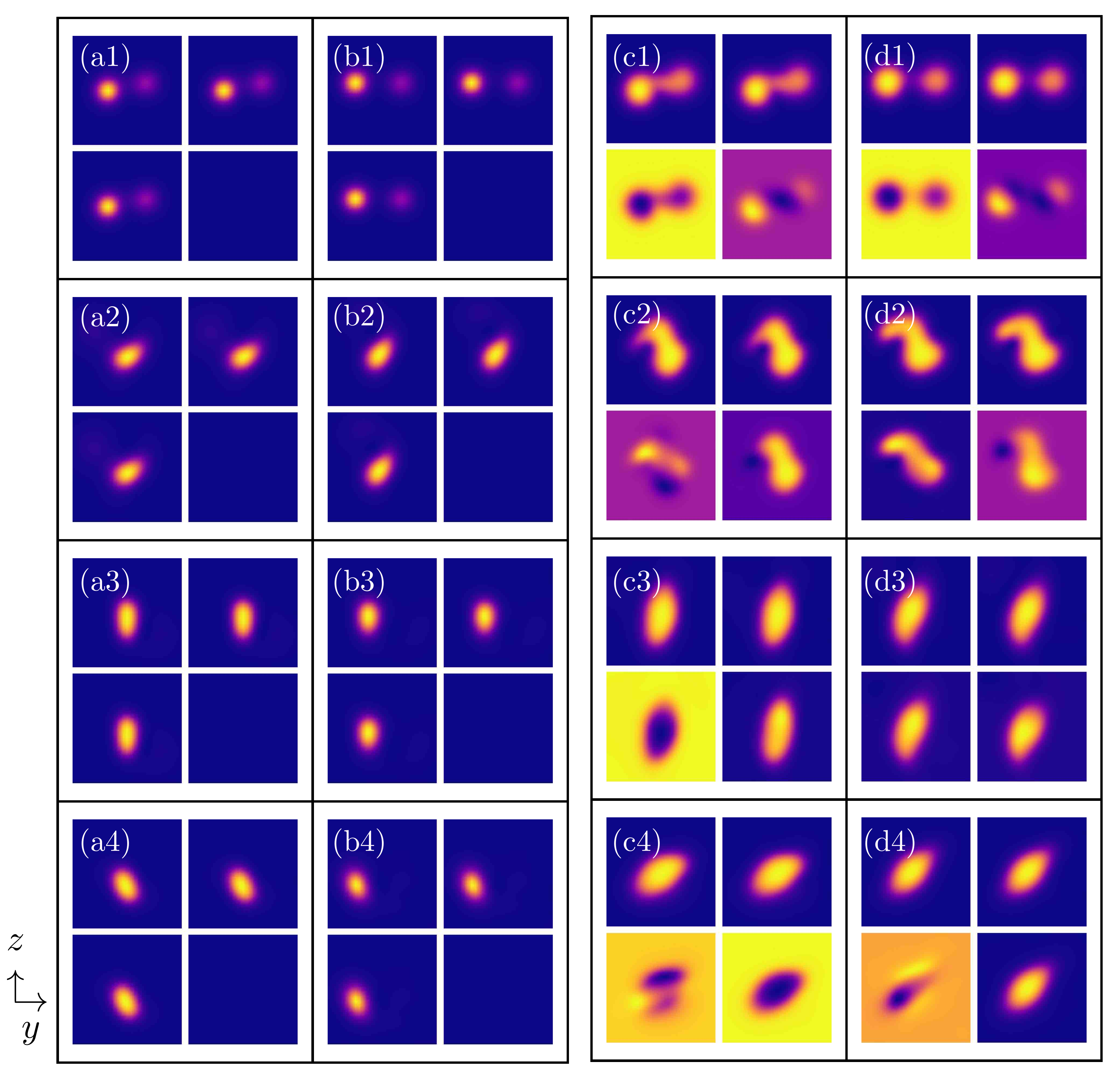}
\caption{Comparative illustrations  of the  evolution of the (odd rows) atomic density and of the (even rows) fidelity of the (a,b) homonuclear and (c,d) heteronuclear (left) real part and (right) imaginary part of two quantum droplets in a binary collision.  
The~adjacent columns of the density for a given letter correspond to different atomic species. The~numbering (e.g., \textbf{a1},\textbf{a2},\textbf{a3}, and \textbf{a4}) indicates different equidistant times. The~adjacent columns of the fidelity density refer to the real and imaginary parts as defined by Equation~(\ref{eq:F}). In~the homonuclear case, $N_1^{(K)}$ = 79,979, and $N_2^{(K)}$ = 110,011 for the second one.
 In the heteronuclear case, $N_1^{(Rb)}$ = 70,227, $N_2^{(Rb)}$ = 96,818, and $N_i^{(Rb)}/N_i^{(K)}= 1.15$ for the second one. The~kinetic parameter is $k_0=11~\mu$m$^{-1}$. The~snapshots correspond to the following evolution times:
 { (a1,b1)}: 6.3662 ms,  {(a2,b2)}: 12.7324 ms,  {(a3,b3)}: 19.0986 ms, and  {(a4,b4)}: 25.4648 ms for the homonuclear panels and
 {(c1,d1)}: 9.5493 ms,  {(c2,d2)}: 15.9155 ms,  {(c3,d3)}: 22.2817 ms, and  {(c4,d4)}: 28.6479 ms for the heteronuclear panels.  The impact parameters for panels (a,c) $b=2$ and for (b,d) $b=3$ in units of  0.855 $\mu$m. All graphs depict the behavior at the $y$-$z$ plane. \label{fig:PanelX_hh}}
\end{figure}

In all the cases that we explored, the~resulting evolution of the atomic densities  
for the different atomic species maintain, within~numerical accuracy, the~same proportionality factor. That is, we numerically found that during the evolution,
\begin{eqnarray}
\rho_a (\vec r, t)&=& \Psi_a^*(\vec r, t)\Psi_a(\vec r, t) = \frac{N^{(a)}}{N^{(b)}}
\Psi_b^*(\vec r, t)\Psi_b(\vec r, t) = \frac{N^{(a)}}{N^{(b)}}\rho_b (\vec r, t)
\\
\Psi_{\alpha}(\vec r, t)&=& \psi_{\alpha_1}(\vec r, t) + \psi_{\alpha_2}(\vec r, t),\quad\alpha = a,b\\
N^{(\alpha)} &=& N^{(\alpha_1)} + N^{(\alpha_2)}
\end{eqnarray}
where $\alpha_k$, $k=1,2$ enumerate the two colliding droplets.
Illustrative examples are depicted in  Figure~\ref{fig:PanelX_hh}.
 
\begin{figure}[h]
 \includegraphics[width=.5\textwidth]{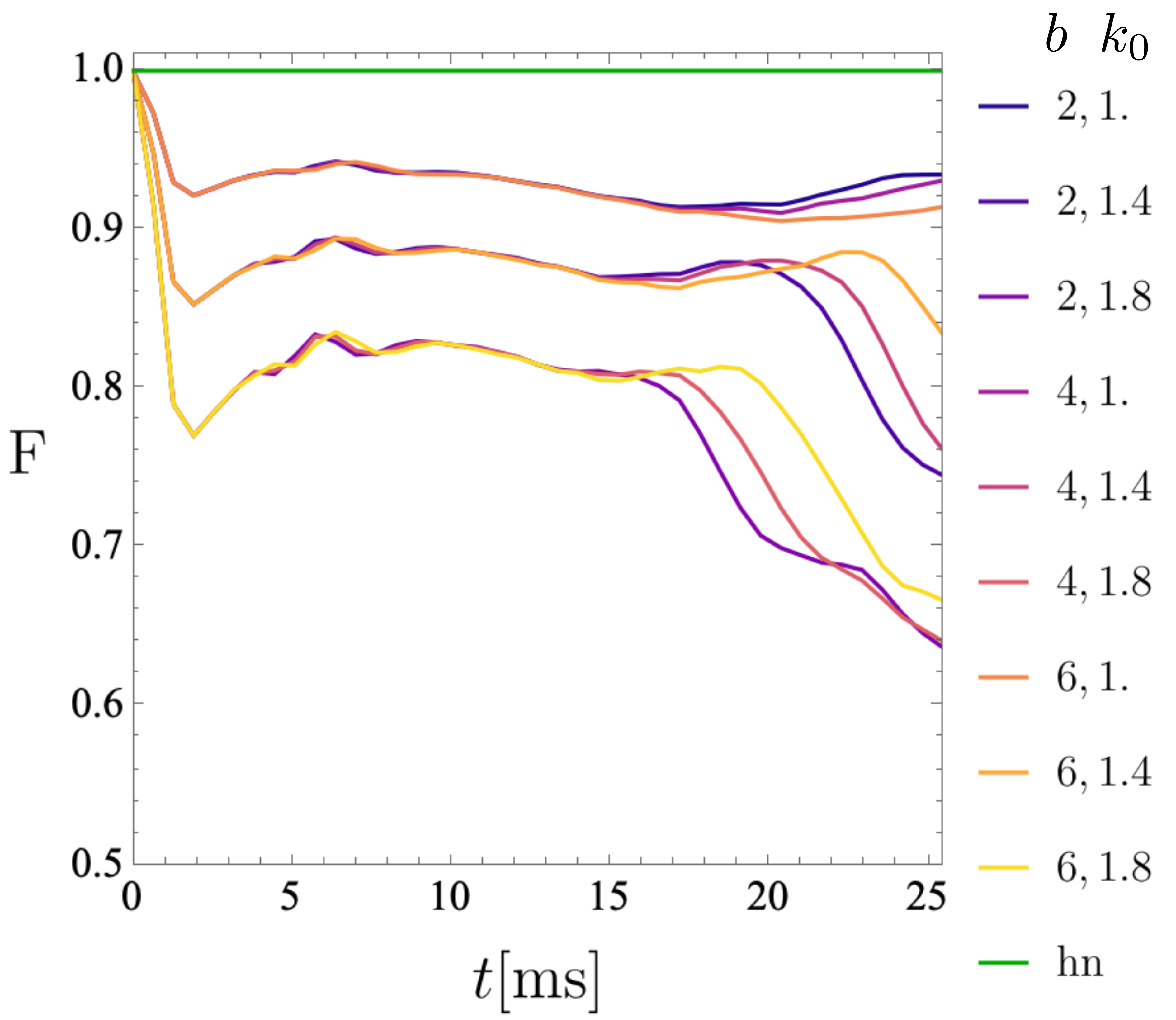}
\caption{Evolution of  the fidelity $F$ for  two heteronuclear quantum droplets in a binary collision. In~this illustrative example, $N_1^{(K)}$ = 70,227, $N_2^{(K)}$ = 96,818, and $N_i^{(Rb)}/N_i^{(K)}= 1.15$.  Different impact parameters $b$ and kinetic factors $k_0$ are considered; the unit of length is 0.855 $\mu$m, so that $k_0$ is given in units of 7.35 $\mu$m$^{-1}$.\label{fig:F_hh} }
\end{figure}

However, when the overlap between the order parameters is estimated using the fidelity $\mathrm{F}$
\begin{equation}
\mathrm{F}(t)=\frac{ \vert\langle\Psi_{a}(t)\vert\Psi_{b}(t)\rangle\vert^2}{\vert\vert\Psi_{a}(t)\vert\vert^2\,\vert\vert\Psi_{b}(t)\vert\vert ^2},\quad \vert\vert\Psi_{a}(t)\vert\vert ^2 = \langle\Psi_{a}(t)\vert\Psi_{a}(t)\rangle
\end{equation} 
as a figure of merit,
it is found to exhibit a nontrivial evolution for heteronuclear droplets and $b\ne 0$, while it equals unity for the collision of homonuclear droplets, as~exemplified in Figure~\ref{fig:F_hh}.
The order parameters are normalized to the total number of atoms in the corresponding species $\vert\vert\Psi_{\alpha}(t)\vert\vert^2 = N^{(\alpha)}$, which in this work we take as constant. This assumption prevents the validity of our calculations for long times compared to those yielding significant atomic
losses~\cite{Semeghini2018, Derrico2019,Ferioli2019,Alba2022}.

A local comparison of the order parameters was also performed in terms of the real and imaginary parts of the fidelity density
\begin{equation}\label{eq:F}
\mathcal{F}(\vec r, t)=\frac{\Psi^*_{a}(\vec r, t)\Psi_{b}(\vec r, t)}{\vert\vert\Psi_{a}(t)\vert\vert \, \vert\vert\Psi_{b}(t)\vert\vert},
\end{equation}  
which is also illustrated in Figure~\ref{fig:PanelX_hh}.  

Since the densities associated with the order parameters $\Psi_{\alpha}(\vec r, t)$ are equal, $\mathrm{F}(t)\ne 1$ implies that different phases were  acquired during the~collision.
\unskip

Since the densities for each atomic species are equal through the collision, the~inertia tensors
\begin{equation}
\bar{I}^{(\alpha)}_{ij}(t) =m_\alpha\int d^3r\Psi^*_{\alpha}(\vec r, t) r_ir_j \Psi_{\alpha}(\vec r, t) 
\end{equation}
differ only by a proportional constant given by the ratio of the total mass of each species in the two droplets: $m_aN^{(a)}/m_bN^{(b)}$.

\begin{figure}[h]
 \includegraphics[width=0.9\textwidth]{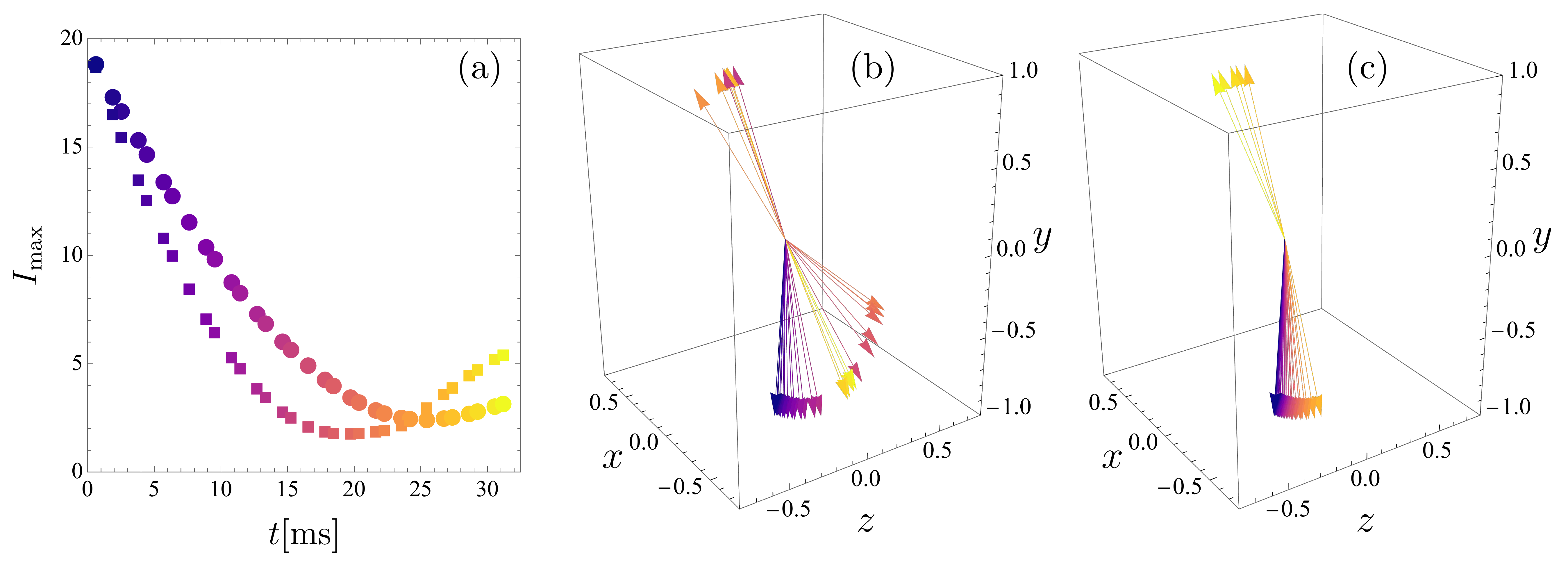}
\caption{Evolution of (\textbf{a}) the largest eigenvalue of the inertia tensor $\bar{I}^{(\alpha)}_{ij}(t)/m_\alpha N^{(\alpha)}$ and
the corresponding eigenvector for (\textbf{b}) a homonuclear ($N_1^{(K)}$ = 79,979  and $N_2^{(K)}$ = 110,011) and (\textbf{c}) a heteronuclear ($N_1^{(Rb)}$ = 70,227, $N_2^{(Rb)}$ = 96,818,  and $N_i^{(Rb)}/N_i^{(K)}= 1.15$) mixture. In~(\textbf{a}), circles correspond to the heteronuclear case and squares to the homonuclear case. The~ unit of length is  0.855 $\mu$m. In~this illustrative example, the two homonuclear droplets 
collide with $b= 3.42~\mu$m and $k_0 =14.7~\mu$m$^{-1}$, while $b=3.42~\mu$m and $k_0 =7.35~\mu$m$^{-1}$ for the heteronuclear case. { The color of each vector in (b) and (c) allows the identification of the evolution time using
the (a) panel.} \label{fig:Imax_hh}}
\end{figure}

The diagonalization of $\bar{I}^{(\alpha)}_{ij}(t)$ allows identification of the principal axes at each time. The~behavior of the direction of the eigenvector corresponding to the largest eigenvalue of the diagonalized $\bar{I}^{(\alpha)}_{ij}(t)$ allows a schematic view of the rotation of the atomic cloud. This is illustrated in Figure~\ref{fig:Imax_hh}. For~simplicity, we report 
$\bar{I}^{(\alpha)}$ divided by the total mass $m_\alpha N^{(\alpha)}$. The~initial value depends directly on the impact parameter. The~long-time evolution 
of the illustrated  eigenvector could also be used to identify  a quasiperiodic global rotation of the coalesced quantum droplet. The~illustrative example in Figure~\ref{fig:Imax_hh} considers the same impact parameter and different kinetic factors $k_0$. For~the same kinetic factors, differences in the evolution of homonuclear and heteronuclear maximum eigenvalues and eigenvectors are still observable but are less evident than expected. 

We also evaluated the evolution of the  linear
\begin{eqnarray}
\vec P &=&-i\hbar\sum_{\alpha=a,b}\int d^3r \Psi_\alpha^\dagger(\vec r) \vec \nabla  \Psi_\alpha(\vec r)\\
&=& N^{(a)}\vec P^{(a)} +N^{(b)} \vec P^{(b)} \\
\vec P^{(\alpha)}& =& -\frac{i\hbar}{N^{(\alpha)}} \int d^3 [(\psi^*_{\alpha_1}(\vec r,t) + 
\psi^*_{\alpha_2}(\vec r,t))\vec \nabla(\psi_{\alpha_1}(\vec r,t) + \psi_{\alpha_2}(\vec r,t))], 
\end{eqnarray}
and angular momenta
\begin{eqnarray}
 \vec L &=&-i\hbar \sum_{\alpha=a,b} 
  \int 
d^3r \Psi_\alpha^\dagger(\vec r)  \vec r \times \vec \nabla  \Psi_\alpha(\vec r)
\\
&=& N^{(a)}\vec L^{(a)} + N^{(b)}\vec L^{(b)} \\
\vec L^{(\alpha)} &=& -\frac{i\hbar}{N^{(\alpha)}} \int d^3 [(\psi^*_{\alpha_1}(\vec r,t) + \psi^*_{\alpha_2}(\vec r,t))\vec r \times\vec \nabla(\psi_{\alpha_1}(\vec r,t) + \psi_{\alpha_2}(\vec r,t))], 
\end{eqnarray}
in terms of the corresponding dynamical densities per number of atoms of each of the two~species.

\begin{figure}[h]
 \includegraphics[width=1\textwidth]{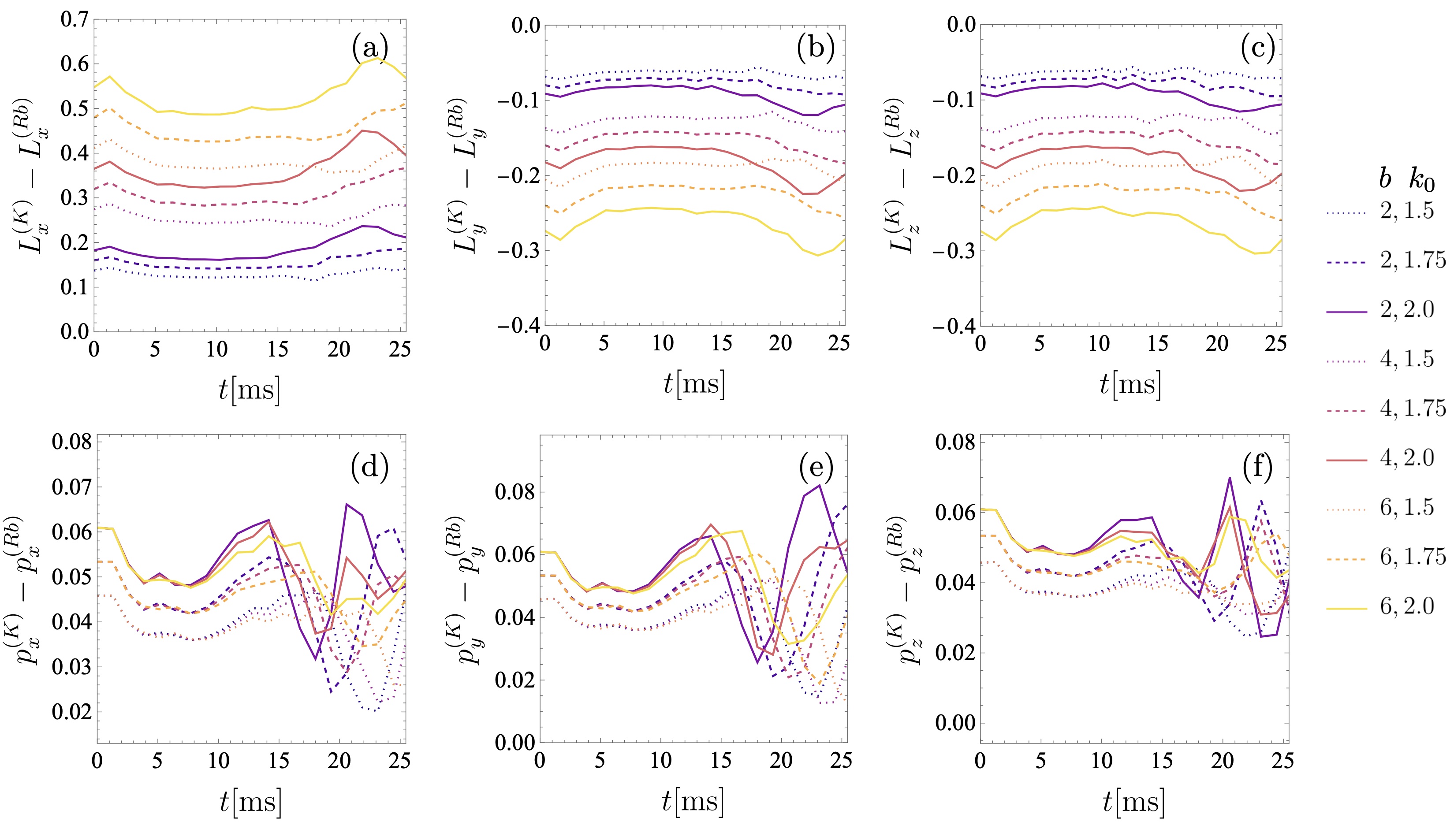}
\caption{ 
Evolution of  the difference of the angular momentum (\textbf{a}-\textbf{c}) and linear (\textbf{d}-\textbf{f}) vector components per atom for  two heteronuclear quantum droplets in a binary collision; $N_1^{(Rb)}$ = 70,227, $N_2^{(Rb)}$ = 96,818, and $N_i^{(Rb)}/N_i^{(K)}= 1.15$. Different impact parameters $b$ and kinetic factors $\hbar k_0$ are considered; the length unit is  0.855 $\mu$m, so that $k_0$ is given in units of 7.35 $\mu$m$^{-1}$.
The angular momentum per atom is given in units of $\hbar$. 
  \label{fig:DLP_hh}
  }
\end{figure}

\begin{figure}[h]
 \includegraphics[width=1\textwidth]{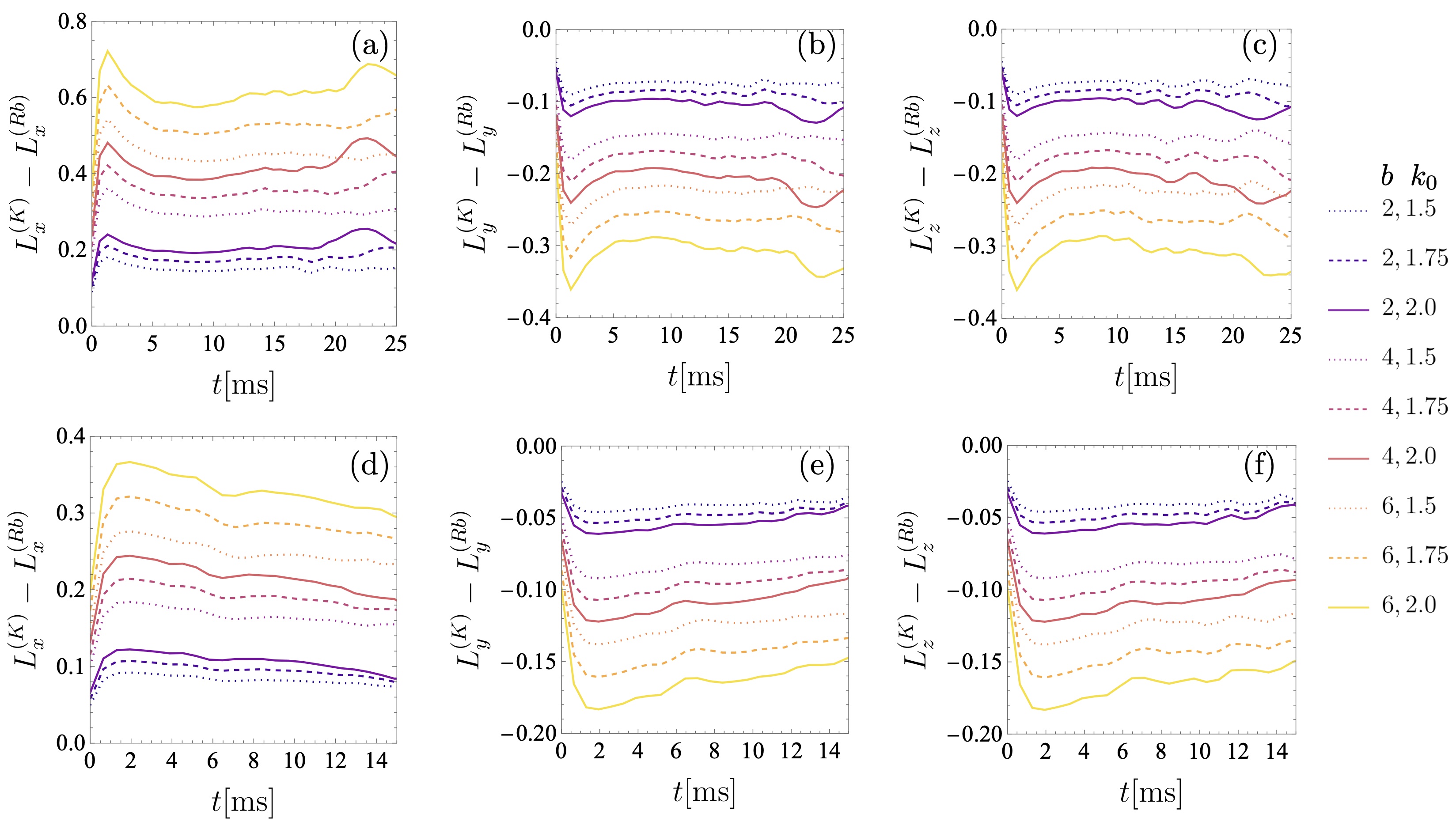}
\caption{
Evolution of  the difference of the angular momentum vector components per atom for  two heteronuclear quantum droplets in a binary collision.
In the first row (\textbf{a}-\textbf{c}), the~droplets are initially identical and $N_1^{(Rb)}=N_2^{(Rb)}$ = 96,818; in~the second row (\textbf{d}-\textbf{f}), $N_1^{(Rb)}$ = 37,000 and $N_2^{(Rb)}$ = 96,818, with~$N_i^{(Rb)}/N_i^{(K)}= 1.15$. The length unit is  0.855 $\mu$m, so that $k_0$ is given in units of 7.35 $\mu$m$^{-1}$. Angular momentum per atom is measured in $\hbar$ units. 
}\label{fig:DLcs_KRb_hh}
\end{figure}

The total linear and angular momentum during the collision is expected---and~numerically confirmed---to~be conserved. For~heteronuclear droplets, the differences in the phases of the order parameters necessarily leads to differences in the linear and angular momentum for each atomic species.
In Figures~\ref{fig:DLP_hh}-\ref{fig:DLcs_KRb_hh},
the~differences are illustrated as a function of the impact parameter $b$ and the kinetic factor $k_0$. As~expected, these differences are greater for smaller values of $b$ and a given $k_0$. The~lightest atoms are more susceptible to the collision~effects.

Up to now, the numerical results explicitly reported in the Figures correspond to binary off-axis collisions of droplets that differ in the number of atoms in each initial droplet, with both being close to the boundary between the incompressible and compressible regimes. In~the case of identical incompressible droplets formed by heteronuclear droplets, the behavior of the difference in the angular momentum for heteronuclear mixtures is similar to that of different incompressible droplets. That is not the case for one droplet in the incompressible regime and the other in the compressible regime. Our numerical simulations indicate that these differences are smaller; the reason can be attributed
to a soft coalescence of the two droplets that allows both species to attain more-similar phases based on their respective order~parameters.

Let us  compare the behavior of the dynamical variables for different collisions. We focus on the
time average of the angular momentum vector, and~we define
\begin{equation}
\Delta L_k (N_1,N_2;N_1^\prime, N_2^\prime)= \frac{\langle L^{(a)}_k(N_1,N_2) - L^{(b)}_k(N_1,N_2)\rangle_t}{\langle L_k^{(a)}(N_1^\prime, N_2^\prime) - L_k^{(b)}(N_1^\prime, N_2^\prime)\rangle_t},\quad k=x,y,z
\end{equation}
 where the pair $(N_1,N_2)$ allows for identification of the quantum droplet ground states in a collision. For~instance, the~collision of two identical droplets corresponds to $N^{(Rb)}_1= N^{(Rb)}_2$. For the heteronuclear droplets under consideration we take $N^{(K)}_1=
N^{(Rb)}_1/1.15$.

We  now compare collisions involving identical droplets, identified by $N_1^{(\alpha)}$, to collisions where the number of atoms $N_2^{(\alpha)}$ in just one of the  droplets has been modified.
For given values of the impact parameter $b$ and the kinetic parameter $k_0$, we
find numerically that
\begin{equation}
\Delta L_k (N_1,N_2;N_1, N_1)= \frac{\langle  L^{(K)}_k(N_1,N_2) - L^{(Rb)}_k(N_1,N_2)\rangle_t}{\langle L^{(K)}_k(N_1, N_1) - L^{(Rb)}_k(N_1, N_1)\rangle_t},\quad k=x,y,z \label{eq:deltaL}
\end{equation}
depends on $N_1$ and $N_2$ but is almost independent of $b$ and $k_0$ within~the range of parameters reported in this work. In~fact, this ratio seems to be simply related to the quotient of the Weber number
\begin{equation}
\Delta L_k (N_1,N_2;N_1, N_1)\approx\frac{(N_2^{(Rb)}+ N_2^{(K)}){\mathrm{We}}(N_2)}{(N_1^{(Rb)} + N_1^{(K)}){\mathrm{We}(N_1)}. } \label{eq:simple}
\end{equation}

\begin{figure}[h]
 \includegraphics[width=1\textwidth]{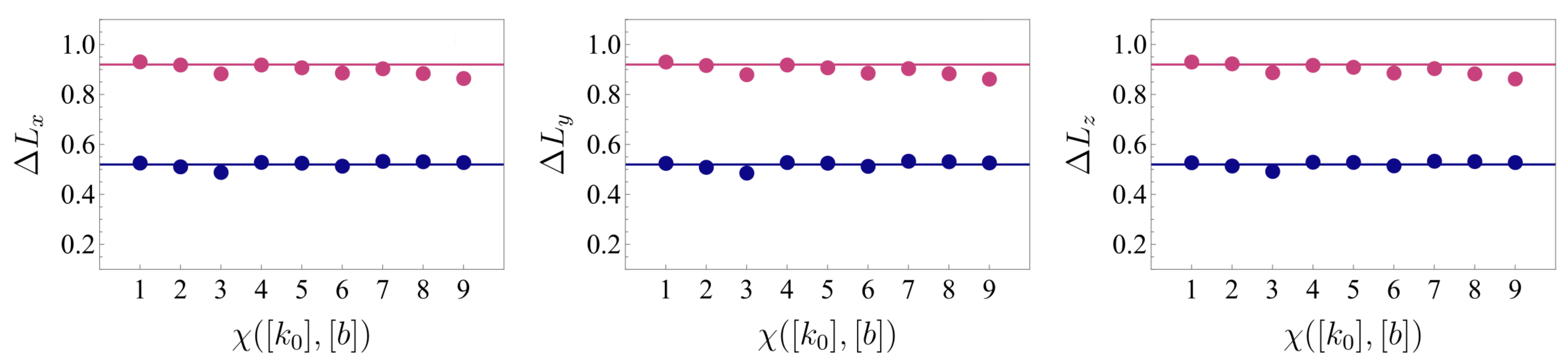}
\caption{
Ratio of the time average of the difference of the angular momentum components  for the two atomic species $\Delta L_k (N_1,N_2;N_1, N_1)$, Equation~(\ref{eq:deltaL}). In~these cases $N_1^{(Rb)}$ = 96,818, $N_2^{(Rb)}$ = 70,227 (magenta circles), and $N_2^{(Rb)}$  = 37,000 (blue circles). The~values of $ [b]=2,4,6$ and $[k_0]=1.5,1.75,2.0$ should be directly substituted in the expression $\chi([k_0],[b])=[b]/2+12([k_0]-3/2)$ to identify the abscissa. 
The~corresponding value of the Weber number ${\mathrm{We}}$ as a function of the number of particles can be evaluated using Table \ref{tab1}.
  \label{fig:DL9_hh}}
\end{figure}

This finding is illustrated in {Figure}~\ref{fig:DL9_hh}. There, we report $\Delta \vec L (N_1,N_2;N_1, N_1)$ for nine pair-collision  
parameters $(b,k_0)$ that were used in the illustrative examples
in Figures~\ref{fig:DLP_hh} and \ref{fig:DLcs_KRb_hh}. The~corresponding value of the Weber number $\mathrm {We}$ as a function of the number of atoms for the ground state was calculated directly from Equations~(\ref{eq:We1}) and (\ref{eq:We2}) and results in the numbers reported in Table \ref{tab1}. The~evolution of the 27 collisions using the EGPE allowed the calculation of the corresponding
difference in angular momenta for the atoms in each species. It was then averaged over time.
This results in a simple relation between both independent calculations, given by Equation~(\ref{eq:simple}).  
{ A clear phenomenological connection between these numerical findings and the dynamics induced by the EGPE remains to be explored.}

The interesting correlation between the differentiated dynamics of each atomic species and the Weber number indicates that---similarly to the classical analogue---the expectations for the dynamical evolution of the collision would be better described by providing both the Weber number (instead of the initial translation energy) and the impact~parameter. 

\begin{table}[h]
\caption{Radius of ground-state droplet $R_0$ for a mixture of Rb and K atoms interacting with the scattering lengths mentioned in the main text. The~scaled Weber number is related to the actual quadrupolar ($\ell =2$) one
by the expression ${\mathrm{We}} = \widetilde{\mathrm{We}}\,k_0^2$. $N_c$ is the critical number of rubidium atoms required to reach the quantum droplet regime for the interaction strengths of the heteronuclear atoms considered in this work. \label{tab1}} 
\begin{tabular}{ccc}
\toprule
 \boldmath{$N^{(Rb)}$}          &   \boldmath{$R_0$ [$\mu$m]} & \boldmath{$\widetilde{\mathrm{We}}$}  		  \\
\midrule 
 96818      = 7.64 $N_c$ &3.36	  &  2.50\\
\midrule
 70227  	   = 5.54 $N_c$	  &2.97 & 3.17\\
\midrule	
 37000  	   = 2.9 $N_c$     &2.31	& 3.40 \\
\bottomrule
\end{tabular}
\end{table}

\section{Conclusions}

Binary collisions of quantum droplets are an ideal scenario for studying the dynamics of quantum degenerate gases without confining potentials. In~this work, we have studied theoretically the behavior of the elementary kinematic and dynamical variables of droplets formed by binary mixtures of homonuclear and heteronuclear droplets.
Low-energy off-axis collisions involve both nontrivial rotational and translational evolution. For~the parameters reported here, the~evolved
dilute degenerate gas seems to result from the coalescence of the initial individual droplets. The { coalesced} droplet rotates as a whole for 
times long enough to admit an experimental corroboration. It is found
that the density of the individual atom species is the same during
the process, while the phase of the corresponding order parameters differs. Direct consequences of this behavior are predicted on the linear and angular momenta; the differences reflected by ct parameter
and seem to be determined by the relation between the translational energy and the surface tension, i.e.,~the Weber number,
in an amazingly simple way that deserves further~analysis.

The experimental observation of the dynamics of quantum droplets before and after a collision is highly constrained by atomic 
losses. Self-evaporation and three-body scattering have been identified as
relevant atomic  
loss mechanisms~\cite{Semeghini2018, Derrico2019,Ferioli2019,Alba2022}.
 In this work, we report calculations before atomic losses are expected to be significant.
The possibility of vortex formation via this excitation method is not excluded, but~given the energies involved in this kind of collision, it seems not to be a dominant effect. Both issues---long-time
 evolution and vortex generation---require further studies involving a larger parameter space. The~possibility of controlling differentiated dynamics of atomic species via binary collisions is an interesting prediction derived from the simulations reported in this work.


\begin{thebibliography}{999}


\bibitem{Petrov2015}Petrov, D.S. Quantum mechanical stabilization of a collapsing Bose-Bose mixture. {\it Phys. Rev. Lett.} {\bf 2015}, {\em 115}, 155302.

\bibitem{Petrov2016} Petrov, D.S.; Astrakharchik, G.E.  Ultradilute low-dimensional liquids. {\it Phys. Rev. Lett.} {\bf 2016}, {\em 117}, 100401.

\bibitem{Cabrera2018} Cabrera, C. R.; Tanzi, L.; Sanz, J.;  Naylor, B.; Thomas, P.; Cheiney, P.;  Tarruell L.   Quantum liquid droplets in a mixture of Bose-Einstein condensates. {\it Science} {\bf 2017}, {\em 359}, {301}.

\bibitem{Semeghini2018} Semeghini, G.; Ferioli, G.; Masi, L.; Mazzinghi, C. ;Wolswijk, L.; Minardi, F.; Modugno, M.; Modugno, G.; Inguscio, M; Fattori, M. Self-bound quantum droplets of atomic mixtures in free space. {\it Phys. Rev. Lett.} {\bf 2018}, {\em 120}, 235301.

\bibitem{Derrico2019}  D'Errico, C.; Burchianti, A.;  Prevedelli, M.;  Salasnich, L.; 
Ancilotto, F.;  Modugno, M.;  Minardi, F.;  Fort, C. Observation of quantum droplets in a heteronuclear bosonic mixture, {\it Phys. Rev. Res.} {\bf 2019}, {\em 1}, 033155.

\bibitem{Guo2021}Guo, Z.;  Jia, F.; Li, L.;  Ma, Y.;   Hutson, J.M.;  Cui, X.;  Wang, D. Lee-Huang-Yang effects in the ultracold mixture of $^{23}$Na and $^{87}$Rb with attractive interspecies interactions, {\it Phys. Rev. Res.} {\bf 2021}, {\em 3}, 033247.

\bibitem{LHY1957}  Lee, T.D.; Huang, K.;  Yang, C.N. Eigenvalues and eigenfunctions of a Bose system of hard spheresand its low-temperature properties. {\it Phys. Rev.}  {\bf 1957}, {\em 106}, 1135.

\bibitem{Larsen1963}  Larsen, D.M.  Binary mixtures of dilute Bose gases with repulsive interactions at low temperature. {\em Ann. Phys.} {\bf 1963}, {\em 24}, {89}.


\bibitem{Ferrier2016} Ferrier-Barbut, I.;  Kadau, H.;  Schmitt, M.;  Wenzel, M.; Pfau, T.  Observation of quantum droplets in a strongly dipolar Bose gas. {\it Phys. Rev. Lett.} {\bf 2016}, {\em 116}, 215301. 

\bibitem{Schmitt2016}  Schmitt, M.;  Wenzel, M.; B\"ottcher, F.; Ferrier-Barbut, I.; Pfau, T.  Self-bound droplets of a dilute magnetic quantum liquid. {\it Nature} {\bf 2016}, {\em 539}, {259}. 
\bibitem{Chomaz2016}  Chomaz, L.;  Baier, S.;  Petter, D.; Mark, M.J.;  W\"achtler, F.; Santos, L.; Ferlaino, F.   Quantum-fluctuation-driven crossover from a dilute Bose-Einstein condensate to a macrodroplet in a dipolar quantum fluid. {\it Phys. Rev. X} {\bf 2016}, {\em 6}, 041039.

\bibitem{Tanzi2019}  Tanzi, L.;  Lucioni, E.; Fam\`a, F.;  Catani, J.;  Fioretti, A.; Gabbanini, C.;  Bisset, R.N.;  Santos, L.;  Modugno, G.  Observation of a dipolar quantum gas with metastable supersolid properties. {\it Phys. Rev. Lett.} {\bf 2019}, {\em 122}, 130405.


\bibitem{Bottcher2019}  B\"ottcher, F.;  Wenzel, M.; Schmidt, J.N.;  Guo, M.;  Langen, T.;  Ferrier-Barbut, I.;   Pfau, T.;  Bomb\'{\i}n, R.;  S\'anchez-Baena, J.;  Boronat, J.; et al. 
Dilute dipolar quantum droplets beyond the extended Gross-Pitaevskii equation. {\it Phys. Rev. Res.} {\bf 2019}, {\em 1}, 033088.


\bibitem{Fetter2009}  Fetter, A.L. Rotating trapped Bose-Einstein condensates. {\it Rev. Mod. Phys. }{\bf 2009}, {\em 81}, 647.

\bibitem{Malomed2019} Malomed, B.A.  Vortex solitons: Old results and new perspectives. {\it Phys. D} {\bf 2019}, {\em 399}, {108}. 

\bibitem{Kartashov2019} Kartashov, Y.V.; Malomed, B.A.;  Torner, L.  Metastability of. Quantum Droplet Clusters. {\it Phys. Rev. Lett.} {\bf 2019}, {\em 122}, 193902.

\bibitem{Luo2021} Luo, Z.-H.;  Pang, W.;  Liu, B.; Li, Y.-Y.; Malomed, B.A.  A new form of liquid matter: Quantum droplets. {\it Front. Phys.} {\bf 2021}, {\em 16}, 32201.

\bibitem{Li2018} Li, Y.; Chen, Z.; Luo, Z.;  Huang, C.;  Tan, H.; Pang, W.; 
Malomed, B.A. Two-dimensional vortex quantum droplets. {\it Phys. Rev. A} {\bf 2018}, {\em 98}, 063602.


\bibitem{Examilioti2020}  Examilioti, P.;  Kavoulakis, G.M. Ground state and rotational properties of two-dimensional self-bound quantum droplets. {\it J. Phys. B} {\bf 2020}, {\em 53}, 175301.

\bibitem{Tengstrand2019} Tengstrand, M.N.; St\"urmer, P.; Karabulut, E.\"O.;  Reimann, S.M.  Rotating binary Bose-Einstein condensates and vortex clusters in quantum droplets. {\it Phys. Rev. Lett.} {\bf 2019}, {\em 123}, 160405.

\bibitem{Sturmer2021}  St\"urmer, P.; Tengstrand, M.N.;  Sachdeva, R.;  Reimann, S.M.
 Breathing mode in two-dimensional binary self-bound Bose-gas droplets. {\it Phys. Rev. A } {\bf 2021}, {\em 103}, 053302.

\bibitem{Kartashov2020}  Kartashov, Y.V.;  Malomed, B.A.;  Torner, T.  Structured heterosymmetric quantum droplets. {\it Phys. Rev. Research} {\bf 2020}, {\em 2}, 033522.

\bibitem{Otajonov2020} Otajonov, S.R.; Tsoy, E.N.; Abdullaev F.K. Variational approximation for two-dimensional quantum droplets. {\it Phys. Rev. E} {\bf 2020}, {\em 102}, 062217.

\bibitem{Kartashov2018}  Kartashov, Y.V.;  Malomed, B.A.;  Tarruell, L.;  Torner, L. Three-dimensional droplets of swirling superfluids. {\it Phys. Rev. A} {\bf 2018}, {\em 98}, 013612.

\bibitem{Ancilotto2018a} Ancilotto, F.;  Barranco, M.; Guilleumas, M.;  Pi, M. Self-bound ultradilute Bose mixtures within local density approximation. {\it Phys.Rev. A} {\bf 2018},  {\em 98}, 053623.

\bibitem{Caldara2022} Caldara, M.;  Ancilotto, F.  Vortices in quantum droplets of heteronuclear Bose mixtures. {\it Phys. Rev. A} {\bf 2022}, {\em 105}, 063328.

\bibitem{Ferioli2019}  Ferioli, G.; Semeghini, G.; Masi, L.;  Giusti, G.; Modugno, G.; 
Inguscio, M.; Gallemi, A,; Recati, A.;  Fattori, M. Collisions of Self-Bound Quantum Droplets. {\it  Phys. Rev. Lett.} {\bf 2019}  {\em 122}, 090401.

\bibitem{Cikojevic2021} Cikojevi\'c, V.;  Vranje\~s Marki\'c, L.;  Pi, M.;  Barranco M.; Ancilotto, F.; Boronat, J.  Dynamics of equilibration and collisions in ultradilute quantum droplets.
{\it Phys. Rev. Res.} {\bf 2021}, {\em 3}, 043139.

\bibitem{Alba2022} Alba-Arroyo, J.E.; Caballero-Benitez, S.F.; J\'auregui, R.  Weber number and the outcome of binary collisions between quantum droplets. {\it  Sci. Rep.} {\bf 2022}, {\em 12}, 18467.

\bibitem{frohn2000} Frohn, A.;  Roth, N.  {\em Dynamics of Droplets}; Springer Science and Business Media:  {Berlin/Heidelberg, Germany}, 2000. 

\bibitem{Cikojevic2018}  Cikojevi\'c, V.; D\~elalija, K.; Stipanovi\'c, P.;  Vranje\~s Marki\'c L.;  Boronat, J. Ultradilute quantum liquid drops. {\it Phys. Rev. B} {\bf 2018}, {\em 97}, 140502.

\bibitem{Hu2020} { Hu, H.; Liu, X.J. Microscopic derivation of the extended Gross-Pitaevskii equation for quantum droplets
in binary Bose mixtures. Phys. Rev. A {\bf 2020}, {\em 102}, 043302. }

\bibitem{Strang1968} Strang, G.  On the construction and comparison of difference schemes. {\it  SIAM J. Numer. Anal.} {\bf 1968}, {\em 5}, {506}. 

\bibitem{Bao2003} Bao, W.; Jaksch, D.;  Markowich, P.A. Numerical solution of the
Gross-Pitaevskii equation for Bose-Einstein condensation. {\it J. Comp. Phys.} {\bf 2003}, {\em 187}, {318}. 

\bibitem{numerical-recipes} Press, W.H. {\em Numerical Recipies in Fortran 77: The Art of Scientific Computing}, 2nd ed.; University of Cambridge: {Cambridge, UK}, 1992. 





\end{thebibliography}
\end{document}